\documentclass[11pt,a4paper]{article}
\pdfoutput=1 

\usepackage{jinstpub}

\title{Unconventional Ideas for Ionization Cooling of Muons}

\author[a]{T.\,L.~Hart,}
\author[a]{J.\,G.~Acosta,}
\author[a]{L.\,M.~Cremaldi,}
\author[b]{D.\,V.~Neuffer,}
\author[a]{S.\,J.~Oliveros,}
\author[b]{D.~Stratakis,}
 \author[a,1]{D.\,J.~Summers%
 \note{Corresponding author.}}
 \author[b]{and K.~Yonehara}
\affiliation[a]{Department of Physics and Astronomy, University of Mississippi-Oxford,\\
University, MS 38677, U.S.A.}
\affiliation[b]{Accelerator Division, Fermi National Accelerator Laboratory,\\
 Batavia, IL 60510, U.S.A.}
\emailAdd{summers@phy.olemiss.edu}

\abstract{Small muon beams increase the luminosity of a muon collider. Reducing 
the momentum and position spreads of muons reduces emittance and leads to small, cool  beams.
Ionization cooling has been observed at the Muon Ionization Cooling Experiment.
6D emittance reduction by a factor of 100,\,000 has been achieved in simulation.
Another factor of 5 in cooling would meet the basic requirements of a high luminosity muon collider.
In this paper we compare, for the first time, the amount of RF needed  in a cooling channel to previous
linacs.
We also outline three methods aimed to help achieve a final factor of 5 in 6D cooling.}


\begin{document}
\maketitle

\section{Introduction}

   Cool muons are needed for small beam sizes  and high luminosity both 
   for a Higgs \hbox{Factory\,\cite{Alexahin1,Alexahin2,Rubbia}}
   and a high energy 
   muon collider\,\cite{Neuffer1994,Ankenbrandt,Palmer2014}.
   Small beam sizes also facilitate the acceleration of  muons for a muon collider\,\cite{Summers}
   or a neutrino factory\,\cite{Albright,IDS}.
   What principles can be used to rapidly create small beams?
   Muons passing through low Z absorbers 
   lose both transverse and longitudinal momentum
   via ionization. 
    RF cavities replace the lost longitudinal momentum.
   The result is transverse cooling from the lowered  transverse momentum.
   Transverse cooling can be exchanged for longitudinal cooling if 
   higher momentum muons  pass through more material.    
    The relations describing  transverse 
     ($\epsilon_{\perp}$ emittance)    
    and longitudinal 
     ($\epsilon_L$ emittance)     
    cooling are given by\,\cite{Neuffer,Neuffer1983}:

	\begin{eqnarray}
		\frac{d\epsilon_{\bot}}{ds} \simeq  -\frac{g_t}{\beta^{2}} \frac{dE_{\mu}}{ds} \frac{\epsilon_{\bot}}{E_{\mu}}  + \frac{1}{\beta^{3}} \frac{\beta^{*}_{\bot}}{2} \frac{(13.6\,  {\rm{MeV}})^{2}}{E_{\mu}m_{\mu} c^{2} L_{R} }
				\end{eqnarray}
		
		\begin{eqnarray}
		\frac{d\epsilon_{L}}{ds} \simeq - \frac{\raisebox{1pt}{g}_{L}}{\beta^{2} E_{\mu}} \frac{dE_{\mu}}{ds} \epsilon_{L}  
		+ \frac{\gamma^{3} \beta_{L}}{\beta c^{2} p^{2}} \pi (r_{e} m_{e}c^{2})^{2} n_{e} 
		(2\! -\! \beta^{2}) 
		\label{eqem} 
	\end{eqnarray}
	where $ dE_{\mu}/ds $ is the  energy loss rate in the material. $ \beta^{*}_{\bot} $ and $ \beta_{L} $ are transverse and longitudinal  
	betatron functions  which  must be made small using strong magnetic and ``RF/low 
	momentum compaction" focusing,  respectively.
	$L_R$ is the absorber radiation length. $(L_R) (dE/ds)$ should  be large to 
	minimize multiple scattering per unit of absorbed energy.
	$g_{L} $ and $g_{t} $ are partition numbers that 
	depend on beam dispersion and absorber geometry and permit 
transverse ($\epsilon_{\perp}$) to longitudinal ($\epsilon_L$)  emittance exchange.

	\begin{eqnarray}
	g_{L,0} = - {\frac{2}{\gamma^{\,2}}} +
	 {\frac{2(\gamma^{\,2} - \beta^2)}
	 {{\gamma^{\,2}}
	\left[ \ln \left[ \frac{{\rule[-3pt]{0pt}{8pt}2 m_{e} c^{2} \gamma^{2}\beta^{2}}} {I(Z)} \right] - \beta^{2} \right] \rule{0pt}{16pt}}	}
	\end{eqnarray}
	
	\noindent
	$\epsilon_{\bot,eq} $ and $\epsilon_{L,eq} $ are the equilibrium emittances which are calculated  as

	\begin{eqnarray}
	\epsilon_{\bot,eq} \simeq \frac{\beta^{*}_{\bot} (13.6\,{\rm MeV})^{2}}{2 g_{t}\beta m_{\mu} c^{2}L_{R} \left( dE/ds \right)} 
	\label{emitT2}
	\end{eqnarray}
	
	\begin{eqnarray}
	\epsilon_{L,eq} \simeq \frac{\beta_{L} m_{e} c^{2} \beta \gamma^{2} (2-\beta^{2}) }{4 g_{L} m_{\mu} c^{2} 
	\left[ \ln \left[ \frac{{\rule[-3pt]{0pt}{8pt}2 m_{e} c^{2} \gamma^{2}\beta^{2}}} {I(Z)} \right] - \beta^{2} \right] }
	\label{emitL2} 
	\end{eqnarray}

The longitudinal betatron function is given by

\begin{eqnarray}
\beta_L = \sqrt{\frac{\lambda_{rf} \, \beta^{\,3} \, \gamma \, m_{\mu} c^{\,2} \,  \alpha_p}{2 \pi \, eV' \cos\phi_s}}
\label{betaL}
\end{eqnarray}

\noindent where  $V'$ is the average RF voltage gradient in a cell,
$\lambda_{rf}$ is the RF wavelength, and $\phi_s$ is the RF phase angle away
from rising zero crossing.
To first order $\alpha_p$ equals $1/\gamma^{\,2}$ for a linac with no dispersion, 
and  |$1/\gamma^{\,2} - 1/\gamma_t^{\,2}|$ for a lattice with dispersion 
such as a ring or a snake.  By making the transition gamma close to the gamma of the beam, the momentum compaction, $\alpha_p$, can be reduced.
This decreases the longitudinal beta function and the longitudinal equilibrium emittance. Running near transition increases longitudinal momentum spread and reduces bunch length
for a given longitudinal emittance. Reducing the RF wavelength does the same thing.  
Neither reduces the transverse partition number which would raise the
transverse equilibrium emittance.  The momentum spread is increased 
by reducing momentum compaction,
which worsens chromaticity and possibly transmission.

\subsection{Observation of  Ionization Cooling at MICE}

The first demonstration of ionization  cooling 
by  the Muon Ionization  Cooling Experiment (MICE) 
shows that  a long proposed method of shrinking a muon beam into a smaller 
volume works.
MICE used
superconducting focus coils surrounding  a liquid
hydrogen absorber 
encased by very thin aluminum windows
to cool millions of  
large emittance
muons. 
The magnets maximize the beam angular spread in the hydrogen.
The angular divergence of a muon beam can be diminished until it reaches equilibrium with 
multiple Coulomb scattering in the material.
So if there is more angular beam spread, there is more room to reduce and  cool the beam
spread.
Ionization energy loss in MICE reduced transverse momentum.
The MICE muon  ionization cooling measurement is  an important  step towards 
a future muon collider, neutrino factory, and other cool muon experiments.

In detail, 
the international Muon Ionization Cooling Experiment 
(MICE\,\cite{Bogomilov2012,Bogomilov2017,Bogomilov2019,Kaplan}) 
at Rutherford Lab in the UK has measured 
transverse ionization cooling\,\cite{MICEcool}
by focusing muons 
with superconducting solenoids
into 
liquid hydrogen\,\cite{Bayliss,Bayliss2} 
and lithium hydride energy absorbers. 
 The technique measures  
an increase in the core density of an ensemble of muons constructed  from 
individual muons\,\cite{Rajaram,Tanaz3,Drielsma}. 
Measurements to improve understanding of muon multiple scattering 
have been performed both at MICE\,\cite{Paolo} 
and at TRIUMF\,\cite{Attwood}.
Data with muons passing through a polyethylene wedge\,\cite{Tanaz2,Tanaz4,Brown} 
have been recorded to allow measurement of longitudinal to transverse 
emittance exchange in both directions\,\cite{Neuffer1998}. 
MICE has a well characterized muon beam\,\cite{Bogomilov2016,Bertoni,Cremaldi}.  

In summary,
MICE has used 140 MeV/c muons with normalized transverse emittances of
6\,mm and 10\,mm.
 A reduction in the transverse amplitude of muons, when they pass through liquid hydrogen and 
 lithium hydride absorbers, is observed.  
No reduction is seen when the absorbers are removed.
The probability that this effect is a fluctuation is less than $10^{-5}$.
Monte Carlo simulations show the same reductions\,\cite{MICEcool2}.

\section{Helical Cooling Channel}

The helical cooling channel has the advantage of using hydrogen which has less multiple scattering than lithium hydride. However, the continuous solenoidal field limits the strength of focusing,
i.e., $1/\beta^*$.  Recently the momentum compaction  of the channel has been lowered by
using additional coils.
This has reduced the longitudinal emittance from 
1540 microns\,\cite{Yoshikawa} to 890 microns\,\cite{Yonehara}.
The hydrogen in the channel neutralizes space charge and permits short bunch
lengths\,\cite{Plasma4,Parkins,Grote}.
Operating RF cavities in magnetic fields can lead to breakdown\,\cite{ Hassanein,Stratakis2010}.
However, tests show that pressurized  hydrogen gas can prevent breakdown in RF cavities
in ionizing beams and magnetic fields\,\cite{Chung}.
Muon transport in plasma is under study\,\cite{Plasma6}.
Using beryllium for RF cavities also ameliorates breakdown in a magnetic field\,\cite{Bowring}.


\section{Rectilinear Cooling Channel}

The rectilinear  ionization cooling channel\,\cite{Balbekov,Stratakis} is a tightly spaced lattice 
containing wedge absorbers for reducing the 
x, y, and z 
momentum of the muon beam, RF cavities 
for restoring longitudinal muon momentum, and solenoids for focusing the beam.
The net loss in transverse momentum results in cooling.
Tilted solenoids 
create dispersion
 spread  the muons in the wedges as a function of momentum.
Higher momentum muons lose more energy.
As a result, emittance is periodically exchanged between the longitudinal and 
transverse degrees of freedom, resulting in cooling in all three phase space planes.

\begin{figure}[h!]
	       	\centering
                     \includegraphics[width=75mm]{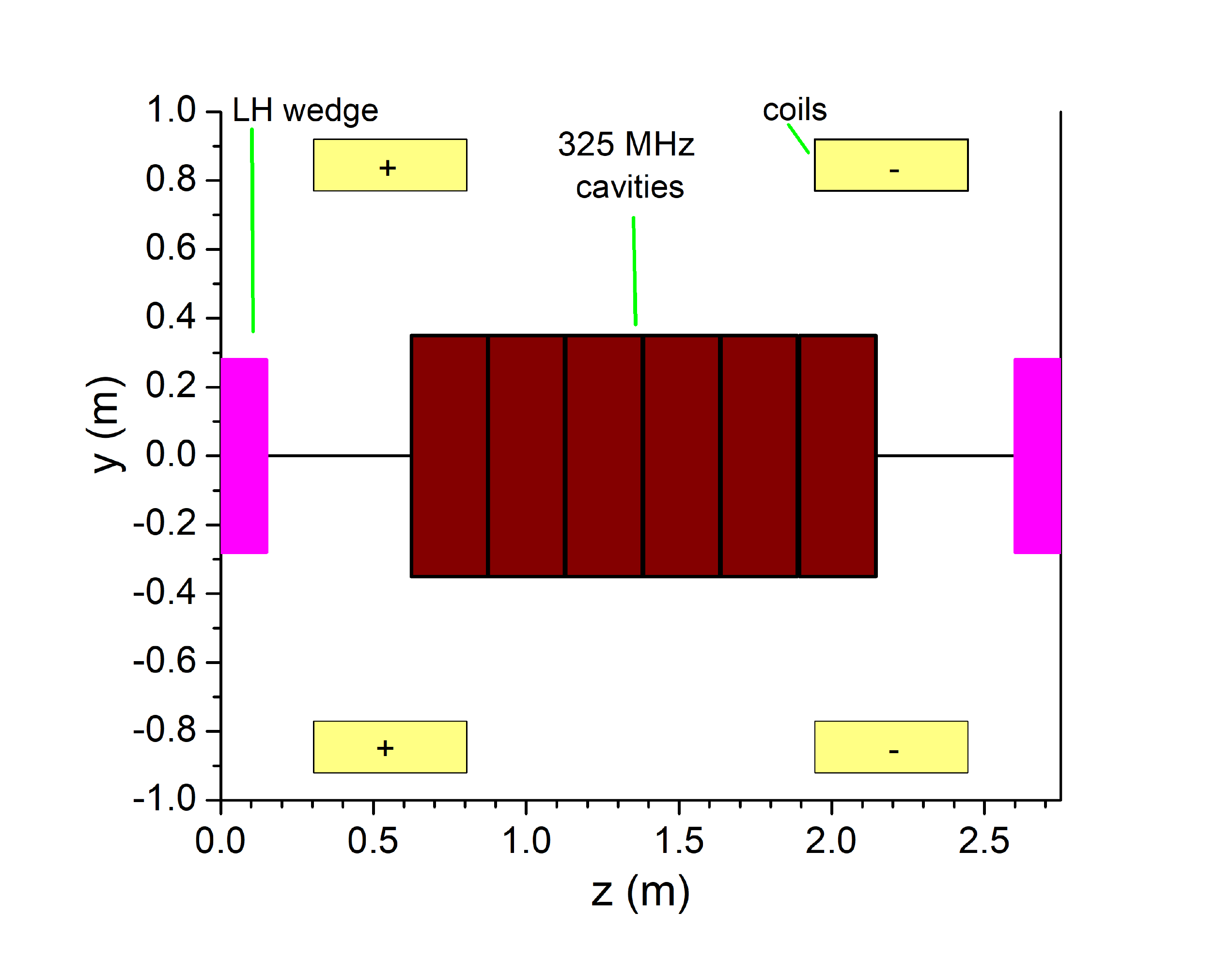}
                        \includegraphics[width=75mm]{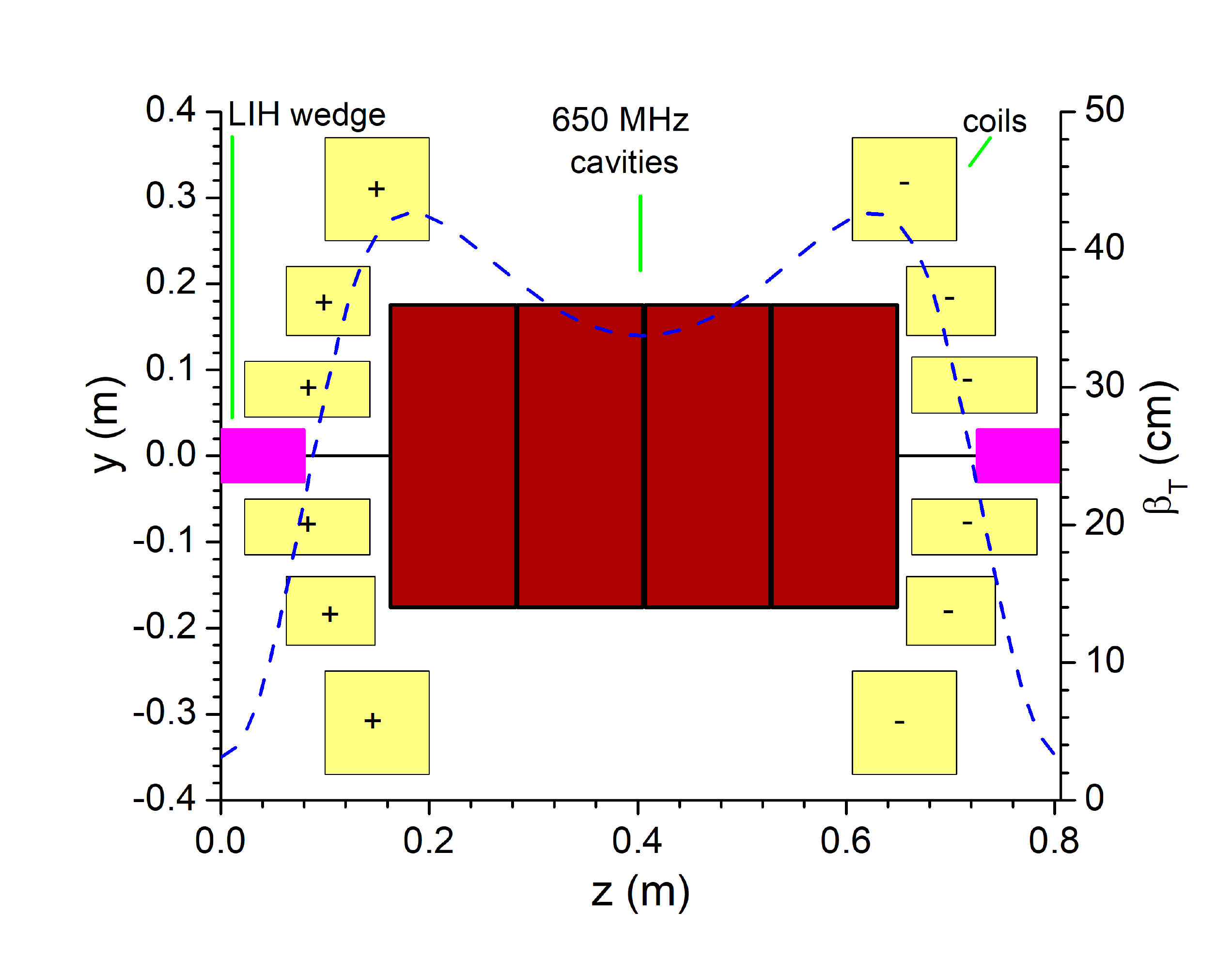}                                       
	       	\caption{One cell of stage B1 in the middle of the Rectilinear Cooling 
		Channel\,\cite{Stratakis} 
		is shown on the left.  B1 follows stages A1, A2, A3, and A4.
		One cell of stage B8 (right) is at the end of rectilinear cooling. }	     	
		 \label{Rect}  
	      \end{figure}

The rectilinear cooling channel has 12 linear stages. 
Cells from stages B1 and B8 are shown in Fig.\,\ref{Rect}.
Stages are constructed  with short  superconducting solenoids up to 14\,T for focusing,  
energy absorbers such as lithium hydride wedges to actually cool muons
by removing transverse momentum,
and 
325 and 650 MHz
normal conducting RF cavities to reaccelerate muons longitudinally.
As one moves along the channel,
each stage  has a smaller minimum 
betatron function to increase cooling and lower emittance.  
Larger maximum betatron functions
in  successive
stages can be tolerated if muons are cooler.
Note that $\beta_{\rm{\,max}} = L^2 / \beta_{\rm{\,min}}$, where $L$ is the distance between the focusing magnets and the 
absorber\,\cite{Sands}.

The rectilinear  channel\,\cite{Stratakis} is tapered 
(also true of the helical channel)
and grows smaller transversely as 
the muon beam cools\,\cite{Stratakis2013}.
Short solenoids provide strong focusing in short regions. 
The solenoids achieve a $\beta^*$ of 3 cm in a lithium hydride absorber
in the twelfth and 
final stage.
In simulation,
the channel  works both with vacuum and with hydrogen to prevent 
breakdown in the RF cavities\,\cite{ Hybrid}.

\subsection{Rectilinear and Helical Cooling Channel Comparison}

Helical\,\cite{Yoshikawa,Yonehara} and rectilinear\,\cite{Stratakis} cooling channels have been simulated as noted in Table \ref{table:channels}. They come close to achieving the cooling desired for a muon collider. The helical channel is based on a gradually tapered 
solenoidal magnetic field with high pressure hydrogen as a continuous absorber.
It achieves a normalized longitudinal emittance of  890 microns\,\cite{Yonehara}, better than the  1570 microns of the rectilinear channel.
Lower longitudinal emittance may also be possible in the rectilinear channel.
The rectilinear channel uses a varying solenoidal magnetic field with lithium hydride wedges
placed in short, low betatron regions for final cooling. It achieves a normalized transverse emittance of 280 microns\,\cite{Stratakis}. 
Noting from eqn.\,\ref{emitT2}  that  $\epsilon_{\bot,eq} \propto   {\beta^{*}_{\bot}}/ {L_{R} \left( dE/ds \right)}$ 
we find that\,\cite{Neuffer}

\begin{eqnarray}
	\frac{\epsilon_{\bot,eq} {\rm(helical)}}
	{\epsilon_{\bot,eq} {\rm(rectilinear)}}} 
	= {\frac{9.5 \, {\rm cm} }  {252.6 \, {\rm MeV}}} \times
	 {\frac{152 \, {\rm MeV} }  {3.0 \, {\rm cm}}	 
	 = 1.9
	\label{ratio}
	\end{eqnarray}

\noindent
The helical channel has less multiple scattering because it uses hydrogen while the rectilinear  channel has a lower 
$\beta^{*}_{\bot}$. 
A 14\,T solenoidal magnetic field gives 
$\beta^{*}_{\bot}$ of 9.5 cm\,\cite{Gallardo}. 
Overall one expects a 1.9$\times$ lower transverse emittance from the rectilinear channel.
This can also be seen in Table \ref{table:channels}.
The helical and rectilinear channels achieve factors of  40,\,000 and
108,\,000 in 6D cooling, respectively.  
Cooling by a factor of 500,\,000  is needed for a $10^{\,34}$ cm$^{-2}$ s$^{-1}$ high energy muon collider. Possible methods to get the
factor of 5 improvement are described in this paper.  The 20\% rectilinear  channel transmission  noted in 
Table \ref{table:RF} can
deliver the $2 \times 10^{12}$ $\mu^+$ and $\mu^-$ bunches needed for the 
$10^{\,34}$ cm$^{-2}$ s$^{-1}$ collider.
There is a factor of 1.5 margin in muon production to allow
for  losses in additional cooling and in acceleration.
 
 \begin{table}[t!]
   \caption{Simulated Helical and  Rectilinear  Cooling Channel  normalized  6D emittances plus the emittance desired for a
high energy   
muon collider in the last line.
The initial 21 bunches are merged during cooling\,\cite{Bao}.}
\vspace{-2mm}
\tabcolsep = 1.5mm
\begin{center}
\renewcommand{\arraystretch}{1.05}
 \begin{tabular}{lcccc} \hline
   & $\epsilon_x$ & $\epsilon_y$ & $\epsilon_z$  & $\epsilon_{6D}$ \\   
   & (mm) & (mm) & (mm)  & (mm$^3$) \\ \hline
   Initial Emittance\,\cite{Stratakis}  & 17.0 & 17.0 & 46.0  & 13,300 \\  
 Helical Cooling 1\,\cite{Yoshikawa}  & 0.523 & 0.523 & 1.54 &   0.421  \\
  Helical Cooling 2\,\cite{Yonehara}  & 0.61 & 0.61 & 0.89 &   0.331  \\ 
 Rectlinear Cooling\,\cite{Stratakis} & 0.28 & 0.28 & 1.57 &   0.123 \\ \hline
 Muon Collider\,\cite{Palmer2007}  & 0.025 & 0.025 & 72 &  0.045  \\ \hline
 \end{tabular}
\end{center}
\label{table:channels}
\end{table}
  

\subsection{Cooling Both Positive and Negative Muons in One Rectilinear Channel}

One 12 stage rectilinear channel (see Table\,\ref{table:RF})
may be able to cool both muon signs, saving a factor of two in cost.  
In each stage, $\mu^+$ and 
$\mu^-$ beams would go in opposite directions, before proceeding to the next stage. Each stage
would require an exit  kicker
on each end for a total of 24 kickers.
The stage A1 exit has the largest normalized transverse emittance, 6.28 mm. 
 The head of a 65 ns long 
bunch train takes 440 ns to travel through the 132 m A1 channel.  A  kicker would need
to  rise in 375 ns (440 $-$ 65). A ring kicker designed for a normalized transverse emittance 
of 10 mm, a rise time of 50 ns, and a 30 m circumference muon cooling ring
required 
5700 kV with one loop\,\cite{Palmer2002}.  The voltage is high. 
Even splitting circuits may not help enough.
The voltage
for this exit  kicker may be low enough to work:

\vspace{-5mm}
\begin{eqnarray}
V_{\rm 1\, loop} = 
(5700 {\  \rm kV}) (6.28 {\,\rm mm}  /10 {\, \rm mm}) (50 {\, \rm ns}/375 {\, \rm ns}) = 480 \ {\rm kV}
\end{eqnarray}

\subsection{Feasibility of the  Rectilinear RF System}

We next calculate the peak and average RF power of the rectilinear cooling  channel.
Then we compare these two values to previous linacs to gauge affordability.
The energy stored per meter in a pillbox RF cavity is given by

\vspace{-5mm}

\begin{eqnarray}
U = {\frac{\epsilon}{2}} \,  E_0^2 \, (J_1(2.405))^2 \, \pi R_c^{\,2} =
{\frac{\epsilon}{2}} \,  E_0^2 \, (0.519)^2 \, \pi \bigg({\frac{2.405 \,  c}  {2 \pi f}}\bigg)^2  = 49400 \, \bigg({\frac{E_0}  {f}}\bigg)^2 \ {\rm joules/m}
\label{eqn:RF}
\end{eqnarray}

\noindent 
where $\epsilon = 8.85 \times 10^{-12}$ Farads/m (the permittivity of free space), $J_1$ is a 
Bessel function, and $c$ is the speed of light. Equation \ref{eqn:RF} is used to 
calculate the `RF total [Joules]'
column in Table \ref{table:RF}.

\begin{table}[b!]
 \caption{Rectilinear RF parameters\,\cite{Stratakis}.
 The RF phase is the angle away from rising zero crossing.
}   
\vspace{-4mm}
\tabcolsep = 0.95mm
\begin{center}
\renewcommand{\arraystretch}{1.05}
 \begin{tabular}{lccccccccccc} \hline
             &               &                 &    &   &   RF       &   & RF   &   &   &  & Muon \\ 
            &   Cell          & Total              &  RF     & RF     &   \#       & RF  & total    & RF    & RF  & RF             &  Trans- \\
 Stage            &  length    & length  & freq  &  gradient  & per  & length   & length   
 & total  & total   & phase & mission \\
(Cells)    &    [m]        &  [m]           & [MHz]      &  [MV/m]  & cell  & [cm]     & [m]   & [GV]   & [Joules] & [deg] & [\%]   \\ \hline
A1 (66)        &    2.000          & 132.00            &  325        & 22.0       & 6         & 25.50   &100.98    & 2.22   & 22900   & 14 & 70.6 \\
A2 (130)        &    1.320          & 171.60            &  325        & 22.0       & 4         & 25.00   &130.00    & 2.85    & 29400   & 15 &  87.5 \\
A3  (107)       &    1.000          & 107.00            &  650        & 28.0       & 5         & 13.49   &72.17      & 2.02     & 6620   & 20  & 88.8  \\
A4 (88)        &    0.800          & 70.40              &  650        & 28.0       & 4         & 13.49   &47.48      & 1.33    & 4350   & 16   &  94.6  \\
B1 (20)         &    2.750          & 55.00              &  325        & 19.0       & 6         & 25.00   &30.00      & 0.57    & 5070   & 41 & 89.7  \\
B2 (32)        &    2.000          & 64.00              &  325        & 19.5       & 5         & 24.00   &38.40      & 0.75   &  6830   & 41 & 90.6 \\
B3 (54)        &    1.500          & 81.00              &  325        & 21.0       & 4         & 24.00   &51.84      & 1.09     & 10700   & 39 &  89.2 \\
B4  (50)       &    1.270          & 63.50              &  325        & 22.5       & 3         & 24.00   &36.00      & 0.81    &   8520  & 49  & 89.7  \\
B5  (91)        &    0.806          & 73.35              &  650        & 27.0       & 4         & 12.00   &43.68      & 1.18    &  3720   & 49 &  87.5 \\
B6  (77)       &    0.806          & 62.06              &  650        & 28.5       & 4         & 12.00   &36.96    & 1.05      &3510  & 49   & 88.0  \\
B7 (50)        &    0.806          & 40.30              &  650        & 26.0       & 4         & 12.00   &24.00    & 0.62     &  1900 & 46 &  89.6 \\
B8 (61)        &    0.806          & 49.16              &  650        & 28.0       & 4         & 10.50   & 25.62    & 0.72      &  2350  & 47 & 89.0 \\ \hline
325 Total&             &  567.10                     &               &               & 1562   &     &  386.77   & 8.29  &  83420 \\
650 Total  &             &   402.27                   &               &              &   2003    &     & 249.91    & 6.92  & 22450 \\
Total      &                       & 969.37             &               &     &  3565           &             & 636.68  &  15.22   & 105870  &   &  20.7  \\
 \hline
\end{tabular}
\end{center}
\label{table:RF}
\end{table}

Next calculate the filling time ($\tau$), duty factor (DF) with an 0.5 $\mu$s flat top, peak 
power ($P$), and 
average power for the 325 MHz RF cavities\,\cite{Derun}, (see Table
\ref{table:Klystron}):

\begin{eqnarray}
\tau = {\frac{Q} {\omega}} = {\frac{25000} {2\pi \times 325 \times 10^{\,6}}}  = 12.2\, \mu{\rm{s}}
\end{eqnarray}
\vspace{-8mm}
\begin{eqnarray*}
 {\rm{Pulse \  Length}} =((3 \times \tau) + 0.5)\, \mu{\rm{s}} = 37.2 \,\mu{\rm{s}} 
\end{eqnarray*}
\vspace{-8mm}
\begin{eqnarray*}
{\rm{DF}} = 37.2 \,\mu{\rm{s}}  \times 15\,{\rm{Hz}} = 5.6 \times 10^{\,-4}
\end{eqnarray*}
\vspace{-8mm}
\begin{eqnarray*}
P = \omega{\frac{U} {Q}} = 2\pi \times 325 \times 10^{\,6} \, {\frac{83420} {25000}}
= 6814\,  {\rm{MW}}
\end{eqnarray*}
\vspace{-8mm}
\begin{eqnarray*}
P_{\rm{AVG}} = 6814\, {\rm{MW}} \times 5.6 \times 10^{\,-4} = 3.8\,{\rm{MW}}
\end{eqnarray*}

\noindent
Now repeat the calculation for the 650 MHz RF cavities:

\begin{eqnarray}
\tau = {\frac{Q} {\omega}} = {\frac{20000} {2\pi \times 650 \times 10^{\,6}}}  = 4.90\, \mu{\rm{s}}
\end{eqnarray}
\vspace{-8mm}
\begin{eqnarray*}
{\rm{Pulse \  Length}} =((3 \times \tau) + 0.5)\, \mu{\rm{s}} = 15.2 \,\mu{\rm{s}} 
\end{eqnarray*}
\vspace{-8mm}
\begin{eqnarray*}
{\rm{DF}} = 15.2 \,\mu{\rm{s}}  \times 15\,{\rm{Hz}} = 2.3 \times 10^{\,-4}
\end{eqnarray*}
\vspace{-8mm}
\begin{eqnarray*}
P = \omega{\frac{U} {Q}} = 2\pi \times 650 \times 10^{\,6} \, {\frac{22450} {20000}}
= 4584\,  {\rm{MW}}
\end{eqnarray*}
\vspace{-8mm}
\begin{eqnarray*}
P_{\rm{AVG}} = 4584\, {\rm{MW}} \times 2.3 \times 10^{\,-4} = 1.1\,{\rm{MW}}
\end{eqnarray*}

 SLAC was powered by 245 65\,MW klystrons, which provided a peak power of 15925 MW to the SLED system, which compressed RF pulses.
 Because of efficiency, AC wall power is roughly twice the average power.
 From Table \ref{table:Klystron}, the  ratio of peak power for muon cooling 
 to that  at SLAC 
 is  (6814 + 4584) / 15925 = 0.72.  The average RF power for cooling muons is less than 
 that in accelerators 
 at SLAC, LEP, and LAMPF.
 RF systems as large as that required for muon cooling have been previously built.
 Higher frequency RF may be possible in the later stages of the
 rectilinear channel to lower power consumption and to lower
 longitudinal emittance.
 
 \begin{table}[t!]
 \caption{RF power comparison}
\vspace{-4mm}
\tabcolsep = 1.2mm
\begin{center}
\renewcommand{\arraystretch}{1.05}
 \begin{tabular}{lcccccccccc} \hline
               &                              &     No. of                           &    RF                  & No. of      & Klystron    & Peak                &  Pulse    & Rep   & Average  \\
              &   Frequency          & RF              &  length     & Kly- & peak         & power               &   Length  & Rate   & Power       \\
Machine    &    [MHz]        &       cavities           & [m]          & strons  & [MW]               &  [MW]                & [$\mu$s] & [Hz] &  [MW]  \\ \hline
Rectilinear  \cite{Stratakis}       &    325         & 1562         &  387  & TBD & TBD     & 6814   
& 37.2 & 15 & 3.8    \\
Rectilinear      &    650          & 2003           &  250  & TBD  & TBD     & 4584      & 15.2 & 15 & 1.1       \\ \hline
SLAC  \cite{Borghi,Ross}       &      2856          &     75000           &  2926  & 245 & 65     & 15925       & 3.5  & 120 & 6.7     \\
LEP \cite{Hubner}      &               352          & 1376         &  585     & 40  & 0.6   & 24      & CW  & CW & 24.0      \\
LAMPF\,\cite{LAMPF}            & 201.25  & & & 4   & 2.5    &  10 & 1000  &  120 & 1.2 \\
LAMPF                        & 805      &  & & 44  & 1.2  &   52.8  &  1000 & 120  & 6.3\\ 
 \hline
\end{tabular}
\end{center}
\label{table:Klystron}
\end{table}

 %
 %


\section{Quadrupole  Cooling Channel}

	       \begin{figure}[b!]
	       	\centering
                     \includegraphics[width=150mm]{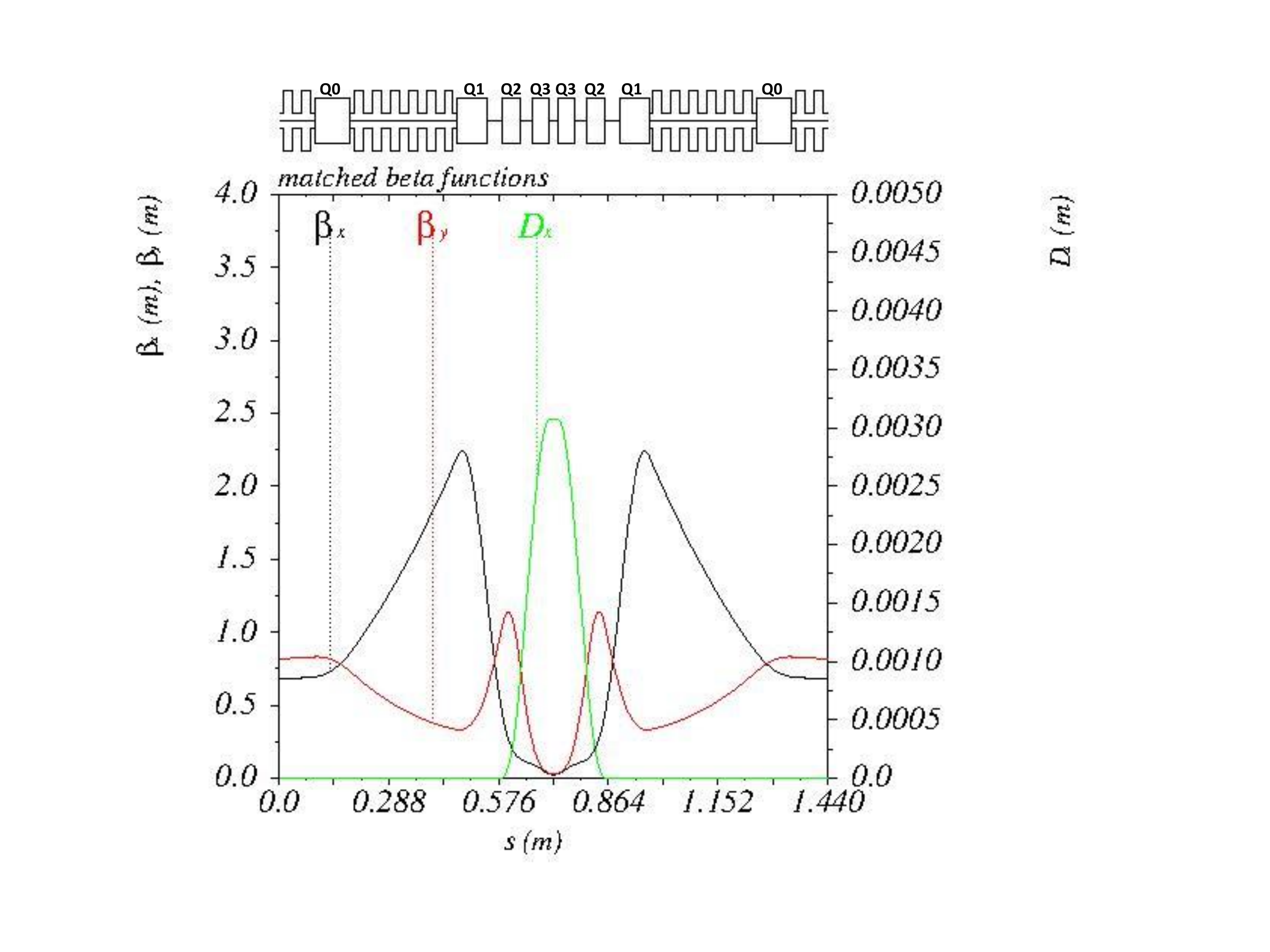}
	       	\caption{Full cell betatron function vs. distance s. The Courant-Snyder\cite{Courant}  parameter evolution through the cell are given by MAD-X   \cite{MAD-X,MAD-X2}.
				The betatron function values are
		 $\mathrm({\beta^{*}_{x},\beta^{*}_{y}})$ = (2.2, 2.7) \,cm in the middle
		 and   $(\beta_x, \, \beta_y)$ =  (0.681, 0.820) m at the ends.}
		     	
		 \label{Beta_madx}  
	      \end{figure}

	 Low equilibrium emittance requires low  $ \mathrm{<\beta_{\bot}>} $. Strong quadrupole focusing\,\cite{Strait,Summers2015,Acosta}  
	 can achieve $ \mathrm{\beta^{*}_{\bot}} $  values 
below the 3 cm achieved  in the final stage of the rectilinear cooling channel design, 
(see Fig.\,\ref{Beta_madx} and Table \ref{table:quads}).
A low $p_z$ spread is used to control chromaticity in the channel.
The input longitudinal emittance is reduced to 632 microns to achieve the low $p_z$ spread.
The longitudinal emittance is reduced by lowering $\beta_L$, which is defined by
eqn.\ref{betaL}.  The 1300 MHz RF frequency helps to do this, as well as the high
RF real estate fraction of 52\%.  There is 0.75 m of RF in a 1.44~m cell.

Reducing beam momentum decreases muon straggling which is good,  but also dictates shorter,
more difficult to build,  quadrupoles. 
A half cell is composed of four quadrupole magnets; the magnet Q0 is a coupling quadrupole preceded by two RF (radio frequency) cavities (L = 0.046875 m) and 
separated from Q1
by six  RF (L = 0.046875 m) cavities.  The RF cavities have a radius of 0.125 m. 
The 1300 MHz RF has a phase angle 11.5$^{\circ}$ away from 
rising zero crossing and a 27.8722 MV/m gradient.

%

		\begin{table}[t!]
   \caption{Superconduction quadrupole parameters.}
\vspace{-2mm}
\tabcolsep = 3.5mm
\begin{center}
\renewcommand{\arraystretch}{1.10}
 \begin{tabular}{ccccc} \hline
      & Bore         & Bore      & Gap to   & Pole Tip\\
     & Diameter   & Length   & Next Quad & Field\\ \hline
 Q0 & 10.0 cm & 9.375 cm & 28.125 cm  & 0.962 T\\
 Q1  & 11.0 cm & 7.875 cm & 3.175 cm & 6.82 T\\
 Q2& 5.3804 cm  & 4.875 cm & 3.0 cm & 9.96 T\\
 Q3 & 3.94832 cm & 4.5 cm & 2.25 cm  & 10.0 T\\ 
 \hline
 \end{tabular}
\end{center}
\label{table:quads}
\end{table}

 \begin{figure}[b!]
 \centering
 	\includegraphics[width=150mm]{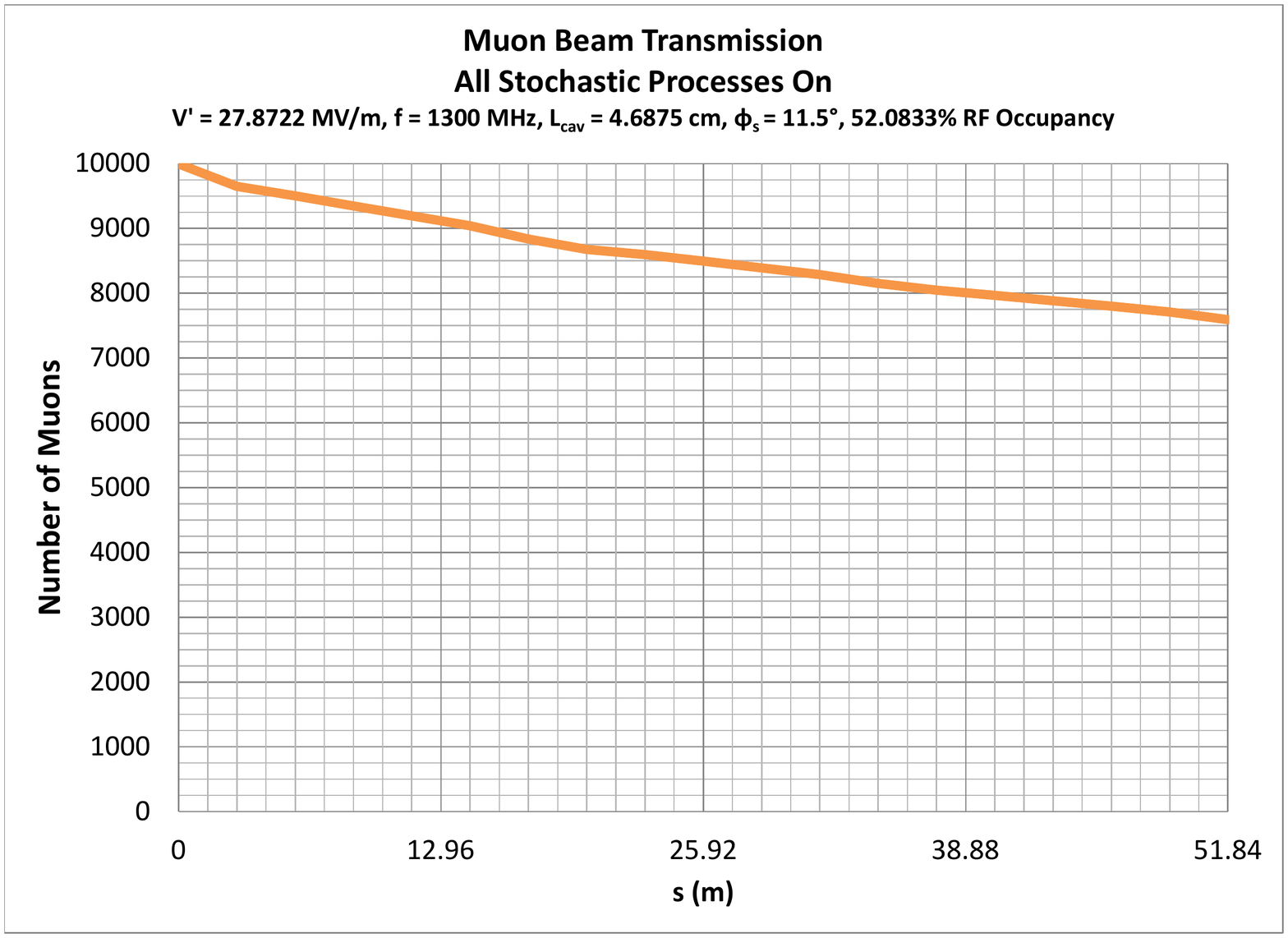} 
 	\caption{76\% transmission through a 51.84 meter long channel (36 full cells).
	At the start of the simulation there are 10000 muons.  At the end of the simulation 
	there are 7600 muons.
	}
 	\label{tran}
 \end{figure}

 The quadrupoles are short and close together.  Fringe fields are an issue\,\cite{Johnstone}.
The quadrupole fields will need to be a 
maximum at the center and then fall\,\cite{Baartman2012}. 
For parameters of the quadrupoles, which use either NbTi or Nb$_3$Sn conductor,
		see Table \ref{table:quads}.
		The Q0 magnet works as a coupling quadrupole reducing the betatron function maximum and allows 
	  the addition of  more RF cavities to increase longitudinal synchrotron focusing.   
	  The quadrupole Q3 is added to reduce both the chromaticity and the minimum beta function. The 144 cm long full cell has a 2.25 cm drift space 
	   for an absorber. The quadrupoles Q2 and Q3 have a dipole magnetic component to produce a uniform dispersion of 2.900 mm at the absorber 
	   space\,\cite{Garren,Garren2,Garren3,Raja}.
	  The betatron function evolution for the full cell is shown in Fig.~\ref{Beta_madx}. The transported beam has $ \beta_{x,max} \cong 2\beta_{y,max}  $.

%

 \subsection{Full Cell Constraints} 
 
  \begin{figure}[b!]
 	\includegraphics[width=150mm]{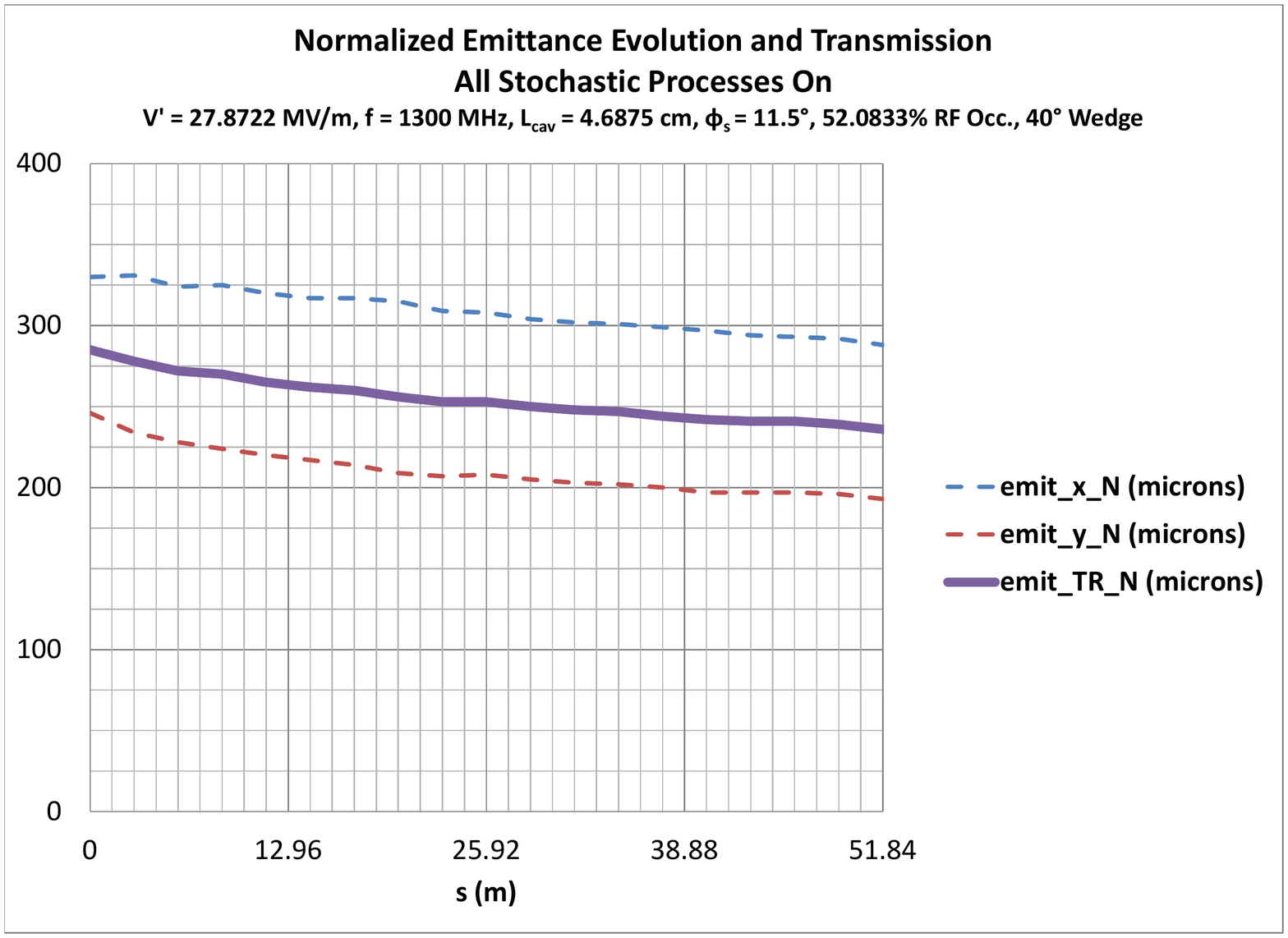}
 	\caption{Transverse emittance evolution for Stage 1. The total 
	transverse emittance goes from $285 \, \mu$m   to $236 \, \mu$m .  
	Normalized emittance, $\epsilon_y$,
	 can be used with the betatron function, $\beta_y$, 
	 to calculate the size of the beam, $\sigma_y$.
	For example,
	$\sigma_y = (\beta_y \, \epsilon_y/ \beta \gamma)^{1/2}$.	\,
	(Emittances are measured
	using  ICOOL's\,\cite{Fernow2000,Fernow2005}	
	ecalcxy tool.)}
 	\label{emitT}
 \end{figure}

 MAD-X \cite{MAD-X,MAD-X2} is used to set magnet parameters to constrain 
 the beta functions and dispersion at the center of the absorber.
Dispersion is flat and constant at the absorber locations 
and zero at the cell ends. The average transverse betatron function 
over the 2.25 cm long absorber area is less than 3.0 cm for a 300 MeV/c muon.
 The quadrupole doublet configuration is designed for a beam near 300 MeV/c.  
 $\mathrm({\beta^{*}_{x},\beta^{*}_{y}})$ equals (2.2, 2.7) \,cm at the centers of the absorbers.
 
 
 As shown in Fig.~\ref{Beta_madx},  $\beta^*$ is small only over a limited longitudinal distance, so the absorber must be dense and short\,\cite{Neuffer1983}. For this configuration, the absorber is 2.25 cm long at the reference orbit,
 which is a good match to $\beta^*_x$ and $\beta^*_y$.

 \subsection{Wedge for Emittance Exchange}
 
 The Q2, Q3 quadrupoles have dipole magnetic field components of 0.667 T and 0.475 T calculated to create a constant $ \mathrm{\eta=2.9 \,mm }$ dispersion at the absorber region as Fig.~\ref{Beta_madx} shows.

    \begin{figure}[b!]
           \includegraphics[width=150mm]{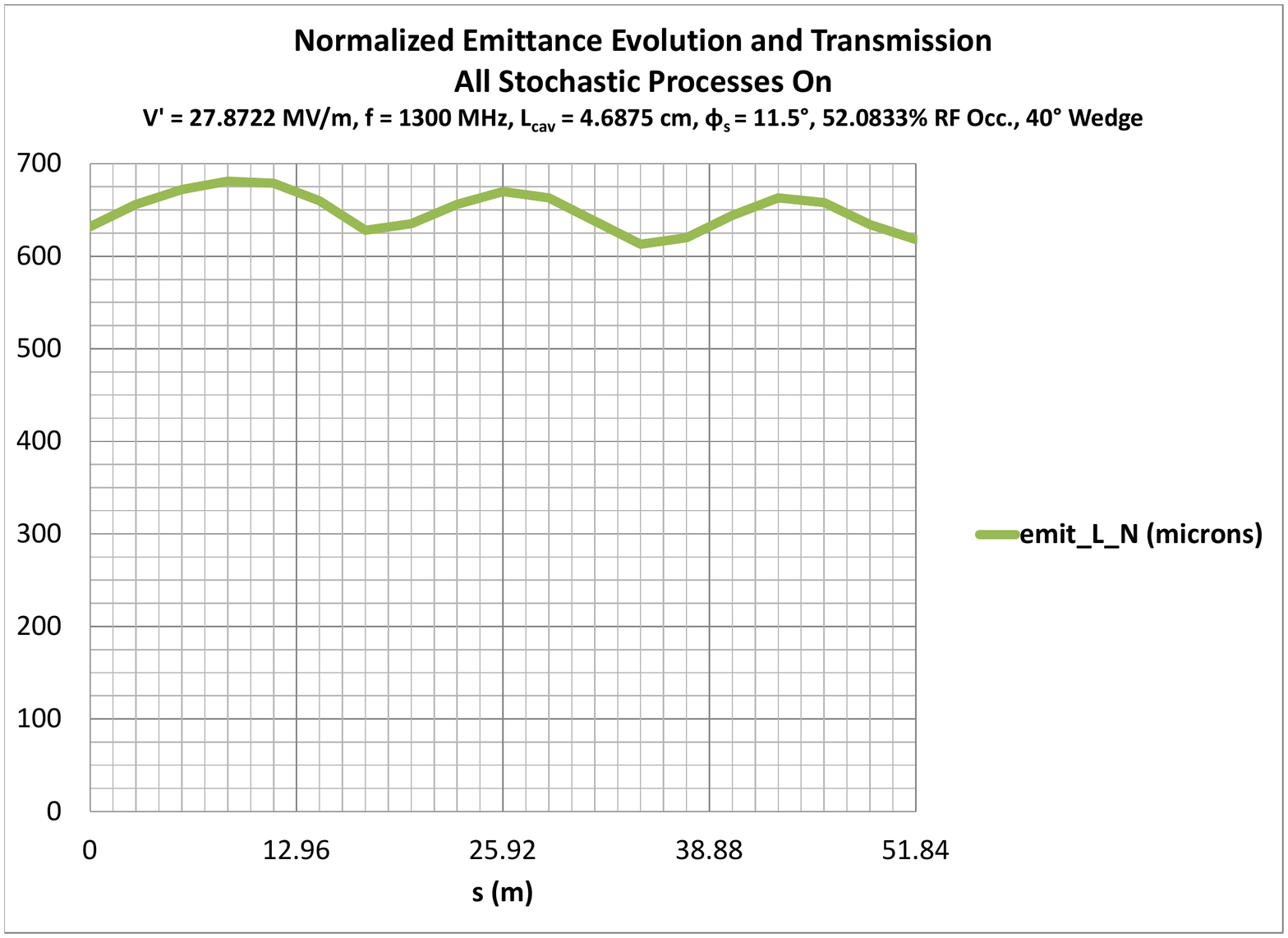}
 	\caption{Longitudinal emittance evolution for Stage 1. The z-x emittance exchange prevents a natural longitudinal emittance increase. 
There are 36 1.44 m cells in the 51.84 m channel.}	
 	\label{emitL}
 \end{figure}

 A  $40^{\circ} $ lithium hydride  wedge  is placed at the center of the 1.44 m long full cell.  The ends of the wedge extend into the Q3 magnet bores on each side of the wedge.
The wedge geometry and the dispersion magnitude modify the partition 
numbers\,\cite{Neuffer}  as follows:
\begin{equation}
g_{L}=g_{L,0}+\frac{\eta}{W}, g_{x}=1-\frac{\eta}{W}
\label{Npartition}
\end{equation}

\noindent
where $ \eta $ is the dispersion magnitude, and $W$ is the distance from the wedge apex to the orbit reference center.  The natural longitudinal partition number, $g_{L,0}$, is $-$0.063; this must be adjusted by emittance exchange between the $(x, z)$ directions through wedge absorbers placed in dispersive regions. 
The values for $ \eta $ and $W$, $\mathrm{2.9 \,mm}$ and $\mathrm{30.909 \,mm}$, respectively, are chosen to reduce the equilibrium longitudinal emittance at the expense of the equilibrium $x$ emittance.  This emittance exchange prevents severe longitudinal beam heating and results in the normalized longitudinal emittance remaining constant, and transverse beam cooling.
Consecutive cells have opposite bending directions to make a snake channel configuration. 



 \subsection{First Stage Simulation}
 

 
 When the absorbers and RF are included and the stochastic processes are enabled, 
the transmission drops to 7600/10000 initial muons. 
Figure \ref{tran} shows transmission vs. distance. 
Higher transmission would be better.  Assume that four stages will be needed.
0.76$^4$ equals 0.33, but 0.90$^{\,4}$ equals 0.65.  Therefore, 90\% transmission requires half as many 
protons to generate muons as 76\% transmission.

 The first channel stage was simulated using ICOOL\,\cite{Fernow2000,Fernow2005}
 and had initial emittances of $ \epsilon_{x,y,z} $ = (0.330, 0.246, 0.632) mm with the three normalized emittances uncorrelated. 
If proper beam  correlations are introduced, transmission improves.
Higher momentum muons need to follow longer path lengths. 

\smallskip
\noindent
 $\epsilon_{x,N,{\rm init}}$ = 0.330~mm is formed with $(\sigma_{x}, \sigma_{p_{x}})_{\rm init}$ = (7.785~mm, 4.412~MeV/c).  
 
 \noindent
 $\epsilon_{y,N,{\rm init}}$ = 0.246~mm is formed with $(\sigma_{y}, \sigma_{p_{y}})_{\rm init} $= (8.147~mm, 3.235~MeV/c). 
 
  \noindent
 $\epsilon_{L,N,{\rm init}}$ = 0.632~mm is formed with $(\sigma_{z}, \sigma_{p_{z}})_{\rm init}$ = (7.4~mm, 8.5~MeV/c) 
 plus an initial correlation between initial $p_{z}$ and initial (radius, $p_{x}/p_{z}$, $p_{y}/p_{z}$) given 
 by: 
 
 \begin{eqnarray}
 \delta p_{z_{\rm{init}}} = 
 (200\, {\rm{MeV/c}})((r_{\rm init}/90.3\,{\rm{mm}})^2+
 (p_{x}/p_{z})_{\rm init}^2+(p_{y}/p_{z})_{\rm init}^2)
 \end{eqnarray}

 \noindent
 When this correlation is added, $p_{z,{\rm mean}}$ increases from 
 300~MeV/c to 303~MeV/c, and $\sigma_{p_{z}}$ increases to 9.05~MeV/c.  
 The added correlation between initial $p_z$ and initial $(r, \, p_x/p_z, \, p_y/p_z)$ 
 makes a small correlation between initial 
$\epsilon_{L{_N}}$ and initial
 $(\epsilon_{x_N}, \, \epsilon_{y_N})$.
  The initial $(r, p_{z})$ correlation improves initial beam matching and transmission through the lattice.


The normalized emittances decrease from
\begin{eqnarray}
 \epsilon_{x,y,z} = (0.330, 0.246, 0.632)\, {\rm mm} \quad {\rm to} \quad
(0.288, 0.193, 0.618) \,{\rm mm}
\label{eqn:emit}
\end{eqnarray}
as Figs.~\ref{emitT} and \ref{emitL} show.
There is a factor of 1.49 6D emittance reduction with 76\% 
transmission through the
first 36 cell stage.
More stages may be possible\,\cite{Summers2015}.  
The transmission needs improvement.
The feasibility of building short, large bore quadrupoles needs to be studied.

In summary,
the quadrupole channel provides a  factor of 1.49 in 6D cooling in 51.84 m.
For comparison the rectilinear channel\,\cite{Stratakis}   cools by
108,000 in 969 m and
1.84 in 51.84 m
(1.84$^{19}$ = 108,000).
One might also make a comparison
to the High Field -- Low Momentum Final Cooling simulations\,\cite{Sayed}.
This 
140\,m long channel has a  solenoidal field of up to 30\,T and a muon momentum 
as low as 70 MeV/c.
Currently the $\epsilon_{xyz}$ normalized emittances change from
(0.300, 0.300, 1.50)\,mm  to
(0.055, 0.055, 76.00)\,mm.
Transverse to longitudinal  emittance exchange works.   
$\epsilon_z$ is fine,  but $\epsilon_x$ and $\epsilon_y$ need to be smaller for a high energy collider.
The 6D emittance rises by a factor of 1.7 
[(0.055\,/\,0.300)\,(0.055\,/\,0.300)\,(76.0\,/\,1.5)].
In the quadrupole scheme, one may be able to  use wedges\,\cite{Neuffer1998}  
or beam slicing\,\cite{Summers2015}  to decrease the final 
transverse emittance at the cost of the longitudinal emittance.
To illustrate,
the normalized longitudinal emittance that can be tolerated by  
a 1.5~TeV/c$^{\,2}$ muon collider final focus  with round 750 GeV/c beams
and a  
$\sigma_p / p = 10^{\,-3}$ chromaticity requirement \cite{Alexahin} is
\begin{equation}
\epsilon_{z} = (\sigma_p /p) \, \Delta z \, (\beta \, \gamma) = 10^{\,-3} \, 10{\rm{mm}} \, 7000 = 70\,{\rm{mm}}
\end{equation}

\noindent
where $\beta \, \gamma$  is the relativistic   factor. 
A final   0.025\,mm normalized transverse emittance
then leads to 
\begin{equation}
 L = \frac{\gamma \, N^{\,2} f_0 \,\,(D\,C)}  {4 \pi  \epsilon_{x,y} \, \beta^{\,*}\rule{0pt}{9pt}}  
 = 
 \frac{7000 \, (2 \times 10^{\,12})^2 \ 180,000/{\rm{s}} \,\,(0.062)} 
  {4 \pi \, (0.0025\,{\rm{cm}}) \, 1.0\,{\rm{cm}}}
= {{1.0 \times 10^{\,\,34}}
{\rm{\,cm}^{\,{-2}} \, s^{-1}}}
\end{equation}

\noindent
where $L$ is average luminosity, $N$ is the initial number of muons per bunch (one positive and one negative),
$f_0$ is the collision frequency,
$D\,C$ is the duty cycle with a 15\,Hz repetition rate, 
and $\beta^{\,*}$ is the betatron function in the collision region.
The beam-beam tune shift, $\xi$, is small enough.

\begin{equation}
\xi = 
\frac{N r_0}  {4 \pi \,\epsilon_{x,y}} =
\frac{ (2 \times 10^{12})\, (1.36 \times 10^{-14} {\rm \, mm})}  {4 \pi \,(0.025 {\, \rm mm)}}
= 0.09
\end{equation}

Four quadrupole channels with progressively lower minimum betatron functions might
be staged to achieve the sought-after factor of five in final cooling ($1.49^{\,4} \approx 5$).
Larger maximum betatron functions in these stages could  be tolerated with cooler muons.
Transmission in  the quadrupole channel is 76\%, which is too low.
Ways  to increase this to 90\% are being sought.
A final betatron 
function of  approximately 1 cm with strong focusing quadrupoles is needed.
Calculations indicate that this requirement is possible\,\cite{Summers2015}.

\section{
Parametric Resonance Ionization Cooling}

Our Muon Accelerator Program (MAP) colleagues  have designed a
Parametric-resonance Ionization Cooling (PIC) channel that strongly focuses a muon beam periodically\,\cite{Derbenev, Morozov}. 
The beam angular spread is maximized  at  foci where energy absorbers are placed.
The angular spread of a muon beam  can be  diminished until it reaches equilibrium with 
multiple Coulomb scattering in the absorber material.
If there is more angular beam spread, there is more room to reduce 
the spread
and  cool the beam.
A parametric resonance is driven by periodic quadrupoles.
The quadrupole  wavelength 
is one quarter of the wavelength of the cooling channel and 
drives beam oscillations.
The resonance causes periodic focal points, where the beam becomes 
progressively
narrower in x and wider in x$^{\,\prime}$ as it passes down the channel. 
Without absorbers the  beam would be unstable. 
Placing thin 
beryllium
energy absorbers at the focal points stabilizes the beam by limiting the beam’s angular 
spread at foci.  
Being a resonant process, focusing can be very strong without using
large magnetic fields.  Stronger focusing leads to cooler muons.

In the latest PIC design, skew quadrupoles are used to couple the transverse 
dimensions\,\cite{Afanasev,Sy}.
Coupling may reduce the number of multipoles required for 
aberration 
correction.
One may just need sextupoles and octupoles but not  decapoles.
Current multipole optimization controls muons up to 82 mrad.
For a high  luminosity,
high energy 
muon collider, the muon beam angular spread in the PIC channel 
needs to be approximately 200 mrad.  The angle increases if the needed output emittance 
is smaller (small  $\beta_{\perp}$ required)
and decreases  if the available input emittance is smaller.
Note that
beam size and angular spreads  equal $\sqrt{\beta_{\perp} \epsilon_{\perp} / (\beta \, \gamma)}$
and
$\sqrt{\epsilon_{\perp} / ( \beta_{\perp}  \,  \beta \, \gamma)}$,
where $\beta_{\perp}$ is the transverse betatron function,
$\epsilon_{\perp}$ is the 
normalized
transverse beam emittance, and $\beta \, \gamma$ 
is the relativistic factor.

\section{Passive Plasma Lens Cooling}

\subsection{Observation of the Focusing of Electrons by a Passive Plasma Lens}

Focusing of electron beams has been predicted\,\cite{Plasma5} and 
observed\,\cite{Plasma,Plasma2} for passive plasma lenses.
Ion beam focusing has also been studied\,\cite{ion}.

\subsection{Calculation of the Focusing of Muons by a Passive Plasma Lens}

A  passive plasma lens is similar to a lithium lens\,\cite{Lithium} except the focusing 
current is in the muon beam rather than 
the lithium.
Both follow eqn.\,\ref{lens}.
Electrons 
from hydrogen or lithium hydride plasma
can either be pulled in (for a $\mu^+$ beam) or pushed out (for a $\mu^-$ beam) to cancel space charge.
This leaves an azimuthal magnetic field to focus in both x and y simultaneously.

We now expand on previous work for  muon cooling with  
plasma lenses\,\cite{Yonehara,Palmer95}.
In eqn.\,\ref{lens},.
the track bending radius is $\rho$ and $B^{\,\prime}$ is the magnetic field gradient.
The longitudinal muon velocity 
combined with
the azimuthal magnetic field
produces a radial focusing force
with the following equation of motion\,\cite{Lithium}:

\begin{eqnarray}
{\frac{d^{\,2} \,r}{ds^{\,2}}} + {\frac{B^{\,\prime} r}  {B \rho}} = 0
\label{lens}
\end{eqnarray}

Take a normalized longitudinal emittance\,\cite{Palmer2000}  of 0.0006 m,  200 MeV/c muons,
$\beta \gamma$ = p\,/\,m = 1.89, and a momentum spread of 10\%:

\begin{eqnarray}
\epsilon_{LN} = \beta \gamma \ {\frac{dp} {p}} \, \sigma_z \qquad
2 \sigma_z = {\frac{2 \,  \epsilon_{LN}} {{\beta \gamma} \ \frac{dp} {p}}} =
\frac{2 (0.0006)} {1.89 \, (0.10)} = 0.0063 {\rm \, m}
\end{eqnarray}

A  bunch length of 0.0063 m with $3 \times 10^{12}$ muons\,\cite{Stratakis} 
and 
$\beta = p/\sqrt{p^2 + m^2\rule{0pt}{10pt}}$ = 0.88
gives a beam current of

\begin{eqnarray}
I = (1.6 \times 10^{-19}) (3 \times 10^{12}) (0.88) (3 \times 10^{\,8} {\rm m/s}  ) / (0.0063 \, {\rm m})
= 20100 \ {\rm A}
\end{eqnarray}

The beam radius at 200 MeV/c for the end of the last stage of the rectilinear channel is:

\begin{eqnarray}
\sqrt{\frac{\beta_x \, \epsilon_x} {\beta \gamma}}  = 
\sqrt{\frac{(30.0\,{\rm mm})(0.280\,{\rm mm})} {1.89}} =   2.1 \ {\rm mm} = 0.0021 \,{\rm m}
\end{eqnarray}

The magnetic field at 0.0021 m with 0.68$^{\,2}$ of the current within 0.0021 m is:
 
\begin{eqnarray}
B = {\frac{\mu_{\,0} \, I \, (0.68)^2} {2 \pi r}} = 0.89 \ {\rm T}
\end{eqnarray}

The field gradient, $B^{\prime}$, is 424 T/m.  
At $p$ equals 0.2 GeV/c, $\beta_{\perp}$ is\,\cite{Lithium}

\begin{eqnarray}
\beta_{\perp} = (B \rho/B^{\,\prime})^{1/2} = (p/(0.3 \,B^{\,\prime}))^{1/2} = 0.040 \ {\rm m}
= 4.0 {\,\rm cm}
\end{eqnarray}

Four centimeters is small enough to combine with and reduce
the 3 cm betatron function in the last stage of the rectilinear channel, which should
increase cooling. 
Note that (10$^{\,9}$ eV/GeV) / (3 x 10$^{\,8}$ m\,/s) = 1/\,0.3
and that $p = 0.3 \,B \rho$.
A lattice with solenoids matching  into a lithium lens has been simulated\,\cite{Fukui}. 
As the beams gets smaller from cooling, the focusing strength increases.
A small leader bunch may help to create the plasma before the main bunch arrives.

\section{Conclusions}

\begin{itemize}

\item
Muons have been cooled using ionization at MICE.  

\item
Cooling by about a factor of  100,\,000 has
been achieved in simulation. 
Another cooling factor of five
is needed to reach a luminosity of 10$^{\,34}$~cm$^{-2}$~s$^{-1}$ for a 
high energy muon collider.

\item
The peak RF power needed by the rectilinear cooling channel
is 72\% of the RF previously installed at SLAC.  Muons of each sign may be able to travel in opposite 
directions though each of the 12 stages of the rectilinear channel,  before proceeding to the 
next stage.  Thus, only one channel would be  needed for both signs.  

\item
Hydrogen
gas has been experimentally
 used to prevent RF breakdown in the presence of ionizing beams and magnetic fields.
The hydrogen may also neutralize space charge permitting shorter bunches and more 
longitudinal cooling and possibly allowing the magnetic field from the beam current to focus the beam itself  for more transverse cooling.  Short bunches increase instantaneous beam current.

\item
Reducing momentum compaction has been shown to decrease longitudinal emittance in the simulation of the latest helical cooling channel\,\cite{Yonehara}   by a factor of 1.7
to 0.890 mm.

\item
Stronger focusing can transversely cool muons beyond the reach of solenoidal channels for a given magnetic field. 6D cooling by a factor of 1.49 in a 51.84 m long strong focusing 
quadrupole
channel is observed in simulation. 
A $y$ emittance of 0.193 mm is achieved, lower than in the helical and rectilinear
channels.

\item
If a cooling combination can  reach
$\epsilon_{x,y,z}(0.190, \,0.190,\,0.700)\, {\rm mm} \to 0.025 \  {\rm mm}^3$
with adequate transmission, cooling is complete for a 
10$^{\,34}$\,cm$^{-2}$\,s$^{-1}$ luminosity, high energy muon collider.
The idea is to lower longitudinal emittance with  reduced momentum compaction
plus higher RF frequency
and then to add a  final low beta stage for more transverse cooling.
\end{itemize}

\acknowledgments
This
work was supported by the 
Fermi Research Alliance, LLC under contract No. DE-AC02-07CH11359 with the 
U.S. Department of Energy through the DOE Muon Accelerator Program (MAP).

\end{document}